\documentclass[twocolumn,aps,showpacs]{revtex4}
\usepackage{color}
\usepackage{graphics}
\usepackage{epsfig}
\usepackage{amsmath}
\usepackage{amssymb}
\usepackage{amsthm}
\usepackage{amsmath,alltt}
\usepackage{graphicx}
\usepackage{amssymb}
\usepackage[latin1]{inputenc}
\usepackage[T1]{fontenc}
\usepackage{color}

\newcommand{\slsh}[1]{{\not \! #1}}

\newcommand{\bea}{\begin{eqnarray}}
\newcommand{\eea}{\end{eqnarray}}

\newcommand{\be}{\begin{equation}}
\newcommand{\ee}{\end{equation}}
\newcommand{\nn}{\nonumber}

\begin{document}
\title{Critical Number of Flavours in QED}

\author{A. Bashir$^1$, C. Calcaneo-Roldan$^2$, L.X. Guti\'errez-Guerrero$^1$ and
M.E. Tejeda-Yeomans$^2$}
\affiliation{
$^1$Instituto de F\'isica y Matem\'aticas, Universidad Michoacana de San Nicol\'as
de Hidalgo, Edificio C-3, Ciudad Universitaria, Morelia, Michoac\'an 58040, M\'exico.\\
$^2$Departamento de F\'isica, Universidad de Sonora, Boulevard Luis Encinas J. y
Rosales, Colonia Centro, Hermosillo, Sonora 83000, M\'exico.\\
}

\begin{abstract}

         We demonstrate that in unquenched quantum electrodynamics (QED),
chiral symmetry breaking ceases to exist above a critical number of fermion flavours $N_f$.
This is a necessary and sufficient consequence of the fact that
there exists a critical
value of electromagnetic coupling $\alpha$ beyond which dynamical mass generation gets triggered.
We employ a  multiplicatively renormalizable photon propagator involving
leading logarithms to all orders in $\alpha$ to illustrate this.
We study the flavour and coupling dependence of the dynamically generated mass
analytically as well as numerically.
We also derive the scaling laws for the dynamical mass as a function of $\alpha$
and $N_f$. Up to a multiplicative constant, these scaling laws are related
through $(\alpha, \alpha_c) \leftrightarrow (1/N_f, 1/N_f^c)$.
Calculation of the mass anomalous dimension $\gamma_m$ shows that it is always
greater than its value in the quenched case. We also evaluate the $\beta$-function.
The criticality plane is drawn in the $(\alpha,N_f)$ phase space which clearly
depicts how larger $N_f$ is required to restore chiral symmetry for an increasing
interaction strength.

\end{abstract}


\maketitle

\date{\today}

The $\beta$-function determines the running of the coupling
constant. In quantum chromodynamics (QCD), its evolution, both in
the ultraviolet and infrared, is crucially influenced by the
number of light quark flavours. Virtual quarks and gluons
contribute to its perturbative tail in a diametrically opposed
manner and the value of $N_f$ determines which would be the
dominant effect. QCD exhibits asymptotic freedom because $N_f$
happens to be less than a critical value of $N_f^{c_1} = 16.5$.
Lattice studies in the infrared indicate that just below this
value, chiral symmetry remains unbroken and colour degrees of
freedom are unconfined~\cite{Appelquist:2009}. Below this
conformal window, for an $8 < N_f^{c_2} < \; \sim 12$ , the
evolution of the beta function in the infrared is such that QCD
enters the phase of dynamical mass generation (DMG) as well as
confinement. QCD is not the only gauge theory where infrared
dynamics responds to the number of fermion flavours in such a
dramatic fashion. It has been established that QED3 possess a
critical number of flavours $N_f^{c_3}$ associated with the
simultaneous emergence of dynamical masses and confinement if the
electron wave function renormalization, photon vacuum polarization
and electron photon vertex are homogeneous functions at the
infrared momenta,~\cite{Bashir:2008,Bashir:2009}.
See~\cite{Miransky:1985} and~\cite{Pisarski:1984} for some
original works on DMG in QCD and QED3, respectively, through the
Schwinger-Dyson equations (SDEs).

One ponders if such criticality also characterizes other gauge
theories in a similar manner. In this paper, we study the flavour
dependence of DMG in QED4 or QED,~\cite{Miransky:1976}. For large
$\alpha$, it is known to exhibit chiral symmetry breaking in the
one loop approximation of the photon
propagator,~\cite{Gusynin:1990}. A consistent solution for coupled
equations for the fermion mass function and photon wave function
renormalization in the bare vertex approximation was obtained
in~\cite{Pennington:1995}.
%
In this article, we demonstrate that the unquenched QED also has a
critical number of flavours $N_f^c$ above which chiral symmetry is
restored. The starting point is the SDE for the electron
propagator
\bea
   S^{-1}(p) = S^{-1}_0(p) + i e^2 \int d^4k \gamma^{\mu} S(k) \Gamma^{\nu}(k,p)
\Delta_{\mu \nu}(q) \;,
\eea
where $q=k-p$, $e$ is the electromagnetic coupling and $S^{-1}_0(p)=\slsh{p}$ is the
inverse bare propagator for massless electrons. We parameterize the full propagator
$S(p)$ in terms of the electron wave function renormalization $F(p^2)$ and the
mass function $M(p^2)$ as~$
   S(p) = {F(p^2)}/{(\slsh{p}-M(p^2))}.$
        $\Delta_{\mu \nu}(q)$ is the full photon propagator which can be conveniently written
as
\bea
\Delta_{\mu \nu}(q) = -\frac{G(q^2)}{q^2}
\left( g_{\mu \nu} - \frac{q_{\mu} q_{\nu}}{q^2} \right)
- \xi \frac{q_{\mu} q_{\nu}}{q^4}  \;,
\eea
where $\xi$ is the covariant gauge parameter such that $\xi=0$ corresponds to the
Landau gauge. $G(q^2)$ is the photon renormalization function.
The full electron photon vertex is represented by $\Gamma^{\mu}(k,p)$.
The form of the
full vertex is tightly constrained by various key properties of the gauge
theory,~\cite{Bashir:2005},
e.g., multiplicative renormalizability of the fermion and the gauge boson
propagators,~\cite{Bashir:1998, Kizilersu:2009}, perturbation theory,~\cite{Bashir:2000},
the requirements of gauge invariance/covariance,~\cite{Ward:1950, Bashir-1:1994, Bashir-2:1996,
Landau:1956, Bashir:LKF-1}
and of course, observed phenomenology,~\cite{Craig:2009}.
The most general decomposition of this
vertex in terms of its longitudinal and transverse components is
\bea
\Gamma^{\mu}(k,p) = \sum_{i=1}^4 \lambda_i (k,p) L_i^{\mu}(k,p) +
\sum_{i=1}^8 \tau_i (k,p) T_i^{\mu}(k,p) \;,  \label{FVertex}
\eea
where $L_1^{\mu}=\gamma^{\mu}$, $L_2^{\mu}=(k+p)^{\mu}(\slsh{k}+\slsh{p})$,
$L_3^{\mu}=(k+p)^{\mu}$ and $L_4^{\mu}=\sigma^{\mu \nu} (k+p)_{\nu}$, where
$\sigma^{\mu \nu}= [\gamma^{\mu}, \gamma^{\nu}]/2$.
The coefficients $\lambda_i$
are determined through the Ward-Takahashi identity relating the electron propagator
with the electron photon vertex~\cite{Ball:1980}~:
\bea
   \lambda_1(k,p)&=&\frac{1}{2} \left[ \frac{1}{F(k^2)} + \frac{1}{F(p^2)} \right] \;, \nn \\
   \lambda_2(k,p)&=&\frac{1}{2}
\frac{1}{k^2-p^2} \left[ \frac{1}{F(k^2)} - \frac{1}{F(p^2)} \right] \;, \nn \\
   \lambda_3(k,p)&=&-\frac{1}{k^2-p^2} \left[ \frac{M(k^2)}{F(k^2)} - \frac{M(p^2)}{F(p^2)} \right] \;
\label{LVertex}
\eea
and $\lambda_4(k,p)=0$.
A simple choice of the transverse coefficients which, combined with the longitudinal component,
renders both $F(p^2)$ the $G(q^2)$
multiplicatively renormalizable for massles fermions, has been constructed
only recently~\cite{Kizilersu:2009}.
The longitudinal coefficient $\lambda_1$ plays a crucial role in ensuring the correct
leading logarithms are summed up for the photon wave function renormalization.
Similarly $\lambda_2$ dictates the multiplicative renormalizability of the photon
propagator. Using this information,
the {\em ansatz} proposed in~\cite{Kizilersu:2009}
makes use of the following four transverse basis vectors as suggested by Ball and Chiu,~\cite{Ball:1980}~:
\bea
  T_2^{\mu}(k,p) &=& p^{\mu} k \cdot q - k^{\mu} p \cdot q  \;, \nn \\
  T_3^{\mu}(k,p) &=& q^2 \gamma^{\mu} - q^{\mu} \slsh{q} \;, \nn \\
  T_6^{\mu}(k,p) &=& - \gamma^{\mu} (k^2-p^2) + (k+p)^{\mu} \slsh{q} \;, \nn \\
  T_8^{\mu}(k,p) &=& - \gamma^{\mu} k^{\lambda} p^{\nu} \sigma_{\lambda \nu}
                     + k^{\mu} \slsh{p} -  p^{\mu} \slsh{k} \;. \label{TBasis}
\eea
\begin{table}
\begin{tabular}{|l|l|l|l|l|l|}
\hline
$N_f$ & $\alpha_c^A$  & $\alpha_c^N$ & $\alpha$ & $N_f^{cA}$ & $N_f^{cN}$ \\
\hline
{\bf 0.5} & 1.69 & 1.7405  & {\bf 2.5} & 1.18 & 1.0253 \\
{\bf 1.0} & 2.27 & 2.4590  & {\bf 3.0} & 1.54 & 1.3205 \\
{\bf 1.5} & 2.94 & 3.3056  & {\bf 3.5} & 1.86 & 1.6209 \\
{\bf 2.0} & 3.74 & 3.9879  & {\bf 4.0} & 2.15 & 2.0123 \\
\hline
\end{tabular}
\caption{We tabulate analytical (indicated with superscript $A$)
and numerical values (indicated with superscript $N$) of $\alpha_c$ for
different values of $N_f$, and $N_f^c$ for
different values of $\alpha$.}
\label{table1}
\end{table}
\noindent The corresponding coefficients are chosen to depend upon $F(p^2)$ in the following
simple manner~:
\bea
\tau_2 &=& \frac{-4}{3(k^4-p^4)}  \left[ \frac{1}{F(k^2)} - \frac{1}{F(p^2)} \right]
-  \frac{2}{3(k^2+p^2)^2} \nn \\
&\times&  \left[ \frac{1}{F(k^2)} + \frac{1}{F(p^2)} \right]
{\rm ln} \left[  \frac{1}{2} \left( \frac{F(q^2)}{F(k^2)} + \frac{F(q^2)}{F(p^2)} \right)  \right] \;, \nn \\
\tau_3 &=& \frac{5}{12(k^2-p^2)}  \left[ \frac{1}{F(k^2)} - \frac{1}{F(p^2)} \right]
+  \frac{1}{3 (k^2+p^2)} \nn \\
&\times&  \left[ \frac{1}{F(k^2)} + \frac{1}{F(p^2)} \right]
{\rm ln} \left[  \frac{1}{2} \left( \frac{F(q^2)}{F(k^2)} + \frac{F(q^2)}{F(p^2)} \right)  \right] \;, \nn \\
\tau_6 &=& \frac{-1}{4 (k^2+p^2)}  \left[ \frac{1}{F(k^2)} - \frac{1}{F(p^2)} \right]  \;,
 \nn \\
\tau_8 &=& 0 \;.  \label{TVertex}
\eea
The coefficient $\lambda_3$, which enters into the description of massive fermions,
is irrelevant to the power law behaviour of both $F(p^2)$ and $G(q^2)$. However,
it is intimately related to the value of the anomalous dimension $\gamma_m$ for the fermion
mass function. In the quenched theory, the ultraviolet behaviour of $M(p^2)$ can be expressed
as
\bea
                 M(p^2) \sim (p^2)^{\gamma_m/2-1}  \label{anomalous}
\eea in the deep Euclidean region. At criticality, the mass
function behaves as Eq.~(\ref{anomalous}) at all momenta. If the
transverse vertex vanishes in the Landau gauge, $\gamma_m=1.058$,
see e.g.~\cite{Gusynin:1994}. However, Holdom and
Mahanta~\cite{Holdom:1988}, using the arguments based on the
Cornwall-Jackiw-Tomboulis (CJT) effective potential technique,
have shown that $\gamma_m$ is strictly equal to 1. The importance
and usefulness of employing the bare vertex was also stressed
in~\cite{Cohen:1989}. If it were true that $\gamma_m=1$, this
would suggest that there is a necessary piece in the transverse
part of the effective vertex which does not vanish in the Landau
gauge. Complete calculation of the fermion-boson vertex at the one
loop in arbitrary gauge and dimensions,~\cite{Davydychev:2000},
reveals that the transverse part of the vertex indeed does not
vanish in the Landau gauge in any space-time dimensions. This fact
may possibly favour Holdom's arguments. Thus it may well be that
the non-zero transverse piece in the Landau gauge cancels out the
$\lambda_3$ piece of the longitudinal component in the equation
for the mass function. Considering this argument, one such vertex
was constructed in~\cite{Bashir-2:1996}. Following suit, consider
the following full vertex \bea
   \Gamma^{\mu}(k,p) &=& \Gamma^{\mu}_{BC}(k,p) + \Gamma^{\mu}_{KP}(k,p) + \Gamma^{\mu}_A(k,p) \;.
\label{ansatz}
\eea
As the subscripts indicate, $\Gamma^{\mu}_{BC}(k,p)$ is the longitudinal Ball-Chiu
vertex, defined by Eq.~(\ref{LVertex}), $\Gamma^{\mu}_{KP}(k,p)$ is the proposal by
Kizilersu and Pennington, Eq.~(\ref{TVertex}), and $\Gamma^{\mu}_A(k,p)$ is the additional
transverse piece which minimally ensures $\gamma_m=1$ in the quenched case.
With this choice of the full vertex, we obtain, in the massless limit
\bea
     F(p^2) = \left( \frac{p^2}{\Lambda^2} \right)^{\nu} \quad
G(q^2) = \left( \frac{q^2}{\Lambda^2} \right)^s  \;,
\label{MRPropagators} \eea where $\nu= \alpha \xi / (4 \pi)$, $s=
\alpha N_f / (3 \pi)$, $\alpha=e^2/(4 \pi)$ and $N_f$ is the
number of massless fermion flavours. All main conclusions are
robust under different truncations, e.g., for 1-loop logarithmic
photon propagator and for a re-summation of the propagators beyond
leading logs.
Near criticality, where the generated masses are small,
one can assume that the power law solutions for the propagators
capture the correct description of chiral symmetry breaking.
We choose to study the resulting equation for the mass function in the convenient Landau gauge.
Results for any other gauge can be derived by applying the Landau-Khalatnikov-Fradkin
transformations~\cite{Bashir:2009, Bashir:LKF-1, Bashir:LKF-2}.
\begin{figure}[t!]
\vspace{0.4cm}
{\centering
\resizebox*{0.45\textwidth}
{0.35\textheight}{\includegraphics{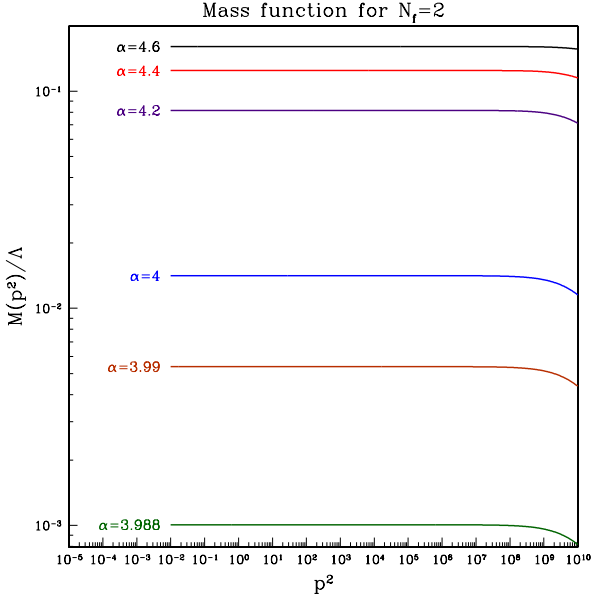}}
\par}
\vspace{-0.3cm}
\caption{The mass function for different values of $\alpha$.}
\label{fig1}
\end{figure}
The usual simplifying assumption $G(q^2)=G(k^2)$ for $k^2 > p^2$
and $G(q^2)=G(p^2)$ for $p^2 > k^2$ allows the analytical treatment
of the linearized  equation for the
mass function~:
\bea
    M(p^2)=  \frac{g(p^2)}{p^2} \hspace{-2mm} \int_{m^2}^{p^2} \hspace{-2mm} dk^2 M(k^2)
   + \int_{p^2}^{\Lambda^2} \hspace{-2mm} dk^2 \frac{M(k^2)}{k^2} g(k^2),
\label{Meq}
\eea
where $g(q^2)= 3\alpha G(q^2)/ (4 \pi)$ and we have introduced ultraviolet cut-off
$\Lambda^2$. The infrared cut-off $m^2$ mimics the $M^2(k^2)$ term in the denominator which
has been dropped off. It is already known that for the one loop photon propagator,
there exists a critical coupling $\alpha_c$ above which masses are dynamically generated.
One can formally demonstrate that the existence of critical coupling implies
the existence of a critical number of flavours above which chiral
symmetry is restored. Note that in Eq.~(\ref{Meq}), $G(q^2)\equiv G(q^2, \alpha N_f)$.
Instead of working with the variables $(\alpha,N_f)$, if we define $\alpha' = \alpha N_f$,
we could equally work with $(\alpha',N_f)$. In such case, $g(q^2)= 3\alpha' G(q^2)/ (4 N_f \pi)$
with $G(q^2) \equiv G(q^2, \alpha')$. If we hold $\alpha'$ constant, the effective coupling
is $1/N_f$. Therefore, the presence of an $\alpha_c$ implies the existence of an $N_f^c$.
The critical behaviour should thus translate as $(\alpha, \alpha_c) \rightarrow (1/N_f,1/N_f^c)$.
Moreover, if there is no critical $\alpha$, there will be no critical $N_f$.
We shall demonstrate this explicitly for the multiplicatively renormalizable photon propagator.
Let us make the change of variables $x=\Lambda^2/p^2$ and convert the integral equation~(\ref{Meq})
into a second order differential equation
\begin{figure}[t!]
\vspace{0.4cm}
{\centering
\resizebox*{0.45\textwidth}
{0.35\textheight}{\includegraphics{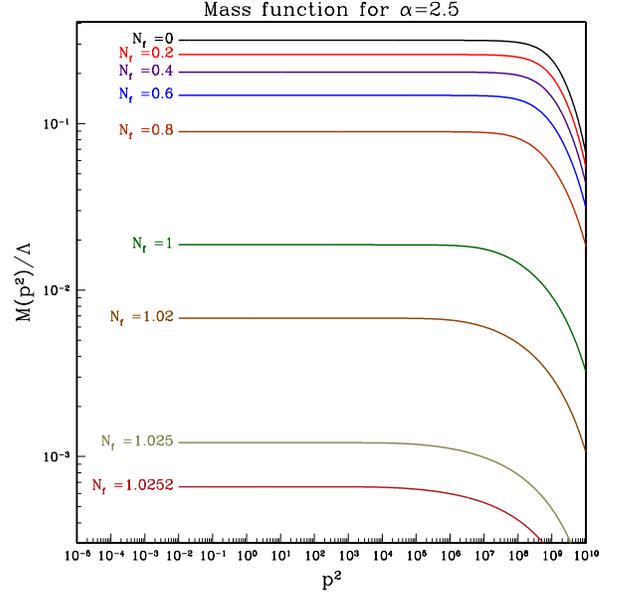}}
\par}
\vspace{-0.3cm}
\caption{The mass function for a fixed $\alpha$ and varying $N_f$.
As $N_f$ is reduced, the mass function drops
significantly for increasingly small variations in $N_f$, suggesting
the existence of a critical number of flavours. }
\label{fig2}
\end{figure}
\bea
     x^2 M''(x) + s x M'(x) + \frac{3 \alpha}{4 \pi} (1-s) \frac{M(x)}{x^s} =0
\label{DE1}
\eea
with the following infrared and ultraviolet boundary conditions respectively
\bea
  M(1) = {M'(1)}/(1-s) \;, \quad M'(\Lambda^2/m^2)=0 \;.
\eea
Eq.~(\ref{DE1}) can be converted into a Bessel equation through Lommel transformations~:
$ z = B x^{\gamma}, \,W = x^{-\lambda} \, M.$
Thus we work with $W(z)$ instead of $M(x)$. The corresponding equation is
\bea
 z^2  W''(z) + z W'(z) + (z^2- A^2) W(z) =0 \;,  \label{DE2}
\eea
where $\gamma=-{s}/{2}$, $\lambda=(1-s)/{2}$, $A= (1-s)/s$ and
$ B= \sqrt{ {3 \alpha} (1-s)/( \pi s^2) }$.
Moreover, we have assumed $s<1$. Eq.~(\ref{DE2}) has the solution
\bea
         W(z) &=& C_1 J_A(z) + C_2 Y_A(z)  \;,  \label{solNf}
\eea
where $J_A(z)$ and $Y_A(z)$ are the Bessel functions of the first and the second kind
respectively.
\begin{figure}[t!]
\vspace{0.4cm}
{\centering
\resizebox*{0.45\textwidth}
{0.35\textheight}{\includegraphics{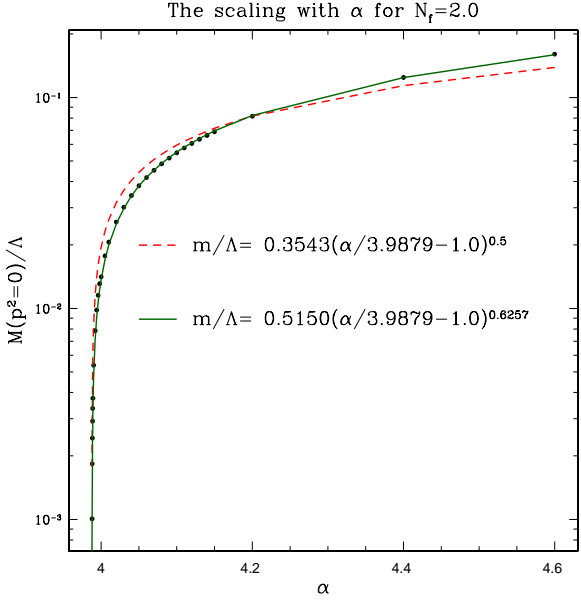}}
\par}
\vspace{-0.3cm}
\caption{The scaling law for the coupling $\alpha$. Numerical solution (solid line)
is compared with analytical prediction (dashed line) of the linearized equation.
 }
\label{fig3}
\end{figure}
The boundary conditions get translated as
\bea
&&      W'(z) = \frac{1-s-\lambda}{\gamma B}\;  W(z) \Bigg|_{z=B}  \;,  \nn \\
&&       \alpha W(z) + \gamma B \left( \frac{\Lambda^2}{m^2}  \right)^{\gamma} W'(z)
= 0 \Bigg|_{z=B  \left( \frac{\Lambda^2}{m^2}  \right)^{\gamma} }  \;.
\eea
These conditions allow us to find the constants $C_1$ and $C_2$, and the equation for
the mass $m$~:
\bea
&& \hspace{-5mm} \frac{ 2 \lambda J_A(z) + \gamma B \left( {\Lambda^2}/{m^2}  \right)^{\gamma}
\left[ J_{A-1}(z)-J_{A+1}(z) \right]
}{ 2 \lambda Y_A(z) + \gamma B \left( {\Lambda^2}/{m^2}  \right)^{\gamma}
\left[ Y_{A-1}(z)-Y_{A+1}(z) \right]  }
\Bigg|_{z=B  \left( \frac{\Lambda^2}{m^2}  \right)^{\gamma}} \nn  \\
&& \hspace{-3mm} = \frac{ \gamma B \left[ J_{A-1}(B)-J_{A+1}(B) \right]-2(1-s-\lambda) J_A(B)}{
\gamma B \left[ Y_{A-1}(B)-Y_{A+1}(B) \right]-2(1-s-\lambda) Y_A(B)} \;.
\label{Keq}
\eea
Critical $\alpha$ or $N_f$ can be obtained from this equation by requiring it
to hold true for $\Lambda \rightarrow \infty$. This implies finding the zeros
of the equation
\bea
   \gamma B \left[ J_{A-1}(B)-J_{A+1}(B) \right]-2(1-s-\lambda) J_A(B) =0.
\eea
For various values of $N_f$, analytical values of $\alpha_c$ have been tabulated in the second column
of~Table.~\ref{table1}. Similarly, for various $\alpha$, $N_f^c$ has been
tabulated in the fifth column of the same table. In order to arrive at the scaling laws,
we expand the left hand side of Eq.~(\ref{Keq}) in powers of $m^2/\Lambda^2$ and keep
the leading terms. Thus
\bea
&& \hspace{-1cm}    \frac{m^2}{\Lambda^2} \equiv f(\alpha,N_f) = \left( \frac{B}{2} \right)^{-2/s}
\Gamma(A) \Gamma(A+2) \frac{(\alpha-\gamma A)}{2 \pi \gamma} \nn \\
&& \hspace{-1cm} \times
 \frac{ \gamma B \left[ J_{A-1}(B)-J_{A+1}(B) \right]-2(1-s-\lambda) J_A(B)}{
\gamma B \left[ Y_{A-1}(B)-Y_{A+1}(B) \right]-2(1-s-\lambda) Y_A(B)} \;.
\eea
Carrying out a Taylor expansion near the critical coupling, we find the following scaling
law
\bea
   {m}/{\Lambda} &=&  h_{\alpha}(N_f)
\; (\alpha - \alpha_c)^{1/2} \;,
\eea
where $ h_{\alpha}(N_f) =
\sqrt{ { \partial f  }/{\partial \alpha} |_{\alpha=\alpha_c} }$.
As anticipated, the scaling law for $N_f$ comes out to be
of the form~:
\bea
   {m}/{\Lambda} &=&  h_{N_f}(\alpha)
\; (N_f^c - N_f)^{1/2} \;,
\eea
\begin{figure}[t!]
\vspace{0.4cm}
{\centering
\resizebox*{0.45\textwidth}
{0.35\textheight}{\includegraphics{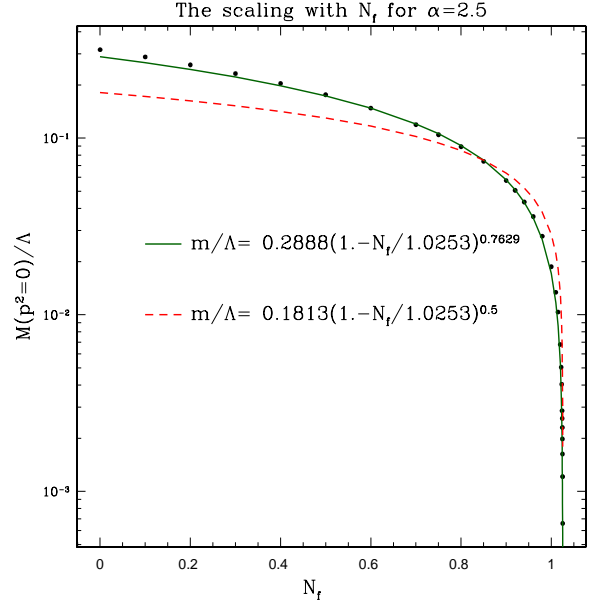}}
\par}
\vspace{-0.3cm}
\caption{The scaling law for the dynamically generated mass as a function
of $N_f$. The solid line is the fit to
the numerical findings whereas the dashed line is the analytical result of
the linearized equation for the mass function.}
\label{fig4}
\end{figure}
\begin{figure}[t!]
\vspace{-1cm}
{\centering
\resizebox*{0.45\textwidth}
{0.3\textheight}{\includegraphics{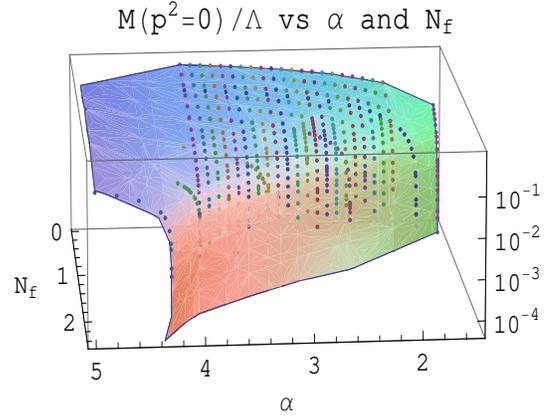}}
\par}
\vspace{-0.6cm}
\caption{The criticality plane for the dynamical mass in the phase
space of $\alpha$ and $N_f$. The points shown are the numerical
results obtained.}
\label{fig5}
\end{figure}
\begin{table}
\begin{tabular}{|l|l|l|l|}
\hline
$N_f$ & $1/b_{\alpha}$  & $\alpha$ & $1/b_{N_f}$ \\
\hline
{\bf 0.5} & 1.0313 & {\bf 2.5} & 0.7629      \\
{\bf 1.0} & 0.7819 & {\bf 3.0} & 0.6944      \\
{\bf 1.5} & 0.6882 & {\bf 3.5} & 0.7314      \\
{\bf 2.0} & 0.6257 & {\bf 4.0} & 0.7640      \\
\hline
\end{tabular}
\caption{The numerical results for the scaling law.}
\label{table2}
\end{table}
\noindent \hspace{-0.4cm} with $ h_{N_f}(\alpha) =  \sqrt{ {
\partial f }/{\partial N_f} |_{N_f=N_f^c}}$. These analytical
results are based upon the linearization of the original problem
and the identification of $M(p^2 \rightarrow 0)=m$. Exact
numerical analysis of the original non-linearized version of
Eq.~(\ref{Meq}) confirms the qualitative nature of the
aforementioned analytical results. In Figs.~(\ref{fig1}) and
(\ref{fig2}), we depict the mass functions for different values of
$\alpha$ and $N_f$. A study of the dependence of $M(0)\equiv m$ as
a function of $\alpha$ and $N_f$, exemplified in
Figs.~(\ref{fig3}) and (\ref{fig4}), permits us to decipher the
corresponding critical values and scaling laws. The numerical
details differ slightly from the analytical findings. We now have
the scaling laws \bea
   {m}/{\Lambda} &=& a_{\alpha}(N_f)  (\alpha - \alpha_c)^{1/b_{\alpha}(N_f) } \;, \\
   {m}/{\Lambda} &=& a_{N_f}(\alpha)  (N_f^c - N_f)^{1/b_{N_f}(\alpha) } \;.
\eea Though none of $b_{\alpha}(N_f)$ and $b_{N_f}(\alpha)$ is
strictly $2$ for a broader fit of the scaling law (see
Table~{\ref{table2}}), figures~(\ref{fig3}) and (\ref{fig4}) show
that the analytical results are not a bad representation of the
exact results in the immediate vicinity of the critical coupling
(compare the solid curves against the dashed ones). The infinite
order phase transition of the quenched QED softens out to a finite
order transition in its unquenched version. It also has
consequences for the mass anomalous dimensions. If  $B \approx C
\equiv |A^2-1/4|$, the large momentum behaviour of
Eq.~(\ref{solNf}) implies $ \gamma_m \approx 1 + {s} $. On the
other hand, if $B >> C$, $\gamma_m \approx 1 + s/2$. The quenched
limit trivially follows. The numerical analysis of the full
equation also yields $\gamma_m > 1$. In order to obtain a finite
electron mass in the limit of $\Lambda \rightarrow \infty$, one
requires charge renormalization {\footnote{Note that QED with
large coupling is not a closed theory. It needs to be supplemented
with certain perturbatively irrelevant operators which become
marginal due to strong QED interactions~\cite{Leung:1986}.}}.
Therefore, in this limit, we impose \bea
   \alpha(\Lambda) &=& \alpha_c + \frac{1}{\left(a_{\alpha}(N_f)
\right)^{b_{\alpha}(N_f)}} \;
\left[ \frac{m}{\Lambda} \right]^{b_{\alpha}(N_f)} \;.
\eea
It implies the following $\beta$-function
\bea
          \beta(\alpha) = \Lambda \; {\partial \alpha}/{\partial \Lambda}
= - b_{\alpha}(N_f) \; (\alpha - \alpha_c) \;.
\eea
Thus the $\beta$-function has a stable zero at the point $\alpha=\alpha_c$, the
result also obtained with the one loop approximation to the photon
propagator,~\cite{Gusynin:1990}.
Not only does a critical value of coupling separate chirally asymmetric and
symmetric phases but so does also a critical number of flavours above which
fermions cease to posses mass just like in QCD and QED3. Fig.~(\ref{fig5})
shows the criticality plane in the phase space of $\alpha$ and $N_f$. It
interpolates and extrapolates the points obtained through the numerical
analysis. We believe that
our analysis can and should be extended to the study of QCD through its
SDEs. This is for the future. \\

We are grateful to R. Delbourgo and V. Gusynin for their comments on the
draft version of this paper.
We acknowledge CIC (UMICH) and CONACyT grants 4.10, 46614-I, 50764 and
94527 ({\em Estancia de Consolidaci\'on}).

\end{document}